\documentclass[10pt,aps,prd,twocolumn,superscriptaddress,amsmath,natbib, nofootinbib]{revtex4-1}

\usepackage{graphicx}
\usepackage{dcolumn}
\usepackage{bm}
\usepackage[mathlines]{lineno}
\usepackage{comment}
\usepackage[T1]{fontenc}
\newcommand\thefont{\expandafter\string\the\font}
\usepackage[hidelinks]{hyperref}
\usepackage[capitalise, nameinlink]{cleveref}
\usepackage{booktabs}
\usepackage{natbib}
\usepackage{cleveref}

\newenvironment{psmallmatrix}
  {\left(\begin{smallmatrix}}
  {\end{smallmatrix}\right)}

\begin{document}

\preprint{APS/123-QED}

\title{Rescuing palindromic universes with improved recombination modeling}

\author{Metha Prathaban}
\affiliation{Astrophysics Group, Cavendish Laboratory, J.J. Thomson Avenue, Cambridge, CB3 0HE, United Kingdom}
\affiliation{Clare College, Trinity Lane, Cambridge, CB2 1TL, United Kingdom}
\author{Will Handley}
\email{wh260@mrao.cam.ac.uk}
\affiliation{Astrophysics Group, Cavendish Laboratory, J.J. Thomson Avenue, Cambridge, CB3 0HE, United Kingdom}
\affiliation{Kavli Institute for Cosmology, Madingley Road, Cambridge, CB3 0HA, United Kingdom}
\affiliation{Gonville \& Caius College, Trinity Street, Cambridge, CB2 1TA, United Kingdom}


\begin{abstract}
\noindent We explore the linearly quantized primordial power spectra associated with palindromic universes. Extending the results of Lasenby \textit{et al.} [Phys. Rev. D \textbf{105}, 083514 (2022)] and Bartlett \textit{et al.} [Phys. Rev. D \textbf{105}, 083515 (2022)], we improve the modeling of recombination and include higher orders in the photonic Boltzmann hierarchy. In so doing, we find that the predicted power spectra become largely consistent with observational data. The improved recombination modeling involves developing further techniques for dealing with the future conformal boundary, by integrating the associated perturbation equations both forwards and backwards in conformal time. The resulting wave vector quantization gives a lowest allowed wave number ${k_0 = 9.93 \times 10^{-5} \textrm{Mpc}^{-1}}$ and linear spacing ${\Delta k = 1.63 \times 10^{-4} \textrm{Mpc}^{-1}}$, providing fits consistent with observational data equivalent in quality to the $\Lambda$CDM model.
\end{abstract}
\maketitle



\section{\label{sec:level1}Introduction}

Astronomical observations~\cite{Perlmutter_1999, Planck2018} have indicated that our current universe is in a state of acceleration, progressing toward an asymptotically de Sitter future; in conformal time, such a universe contains a cosmological coordinate horizon, referred to henceforth as the ``future conformal boundary'' (FCB). Given that this so-called ``end of the universe'' occurs at a finite conformal time, the question thus arises as to what happens to physical quantities such as matter and radiation perturbations at the FCB itself and whether we can continue their development beyond this boundary. In fact, doing so has profound implications and consequences for the observational predictions made by perturbation theory, as demonstrated in previous work by several groups (\citet{Lasenby, Deaglan, Turok}).

\citet{Lasenby} have shown that perturbation variables remain nonsingular at the FCB, and we are able to unambiguously continue them through this boundary. The answer as to what happens to perturbations beyond the FCB lies in considering how they approach the next genuine singularity, the so-called ``big bang 2'' (BB2). Since conformal time forms a ``double cover'' of the solutions, we should demand reflecting boundary conditions. Alternatively, since we are working to linear order in this treatment, we must require that our modes be finite everywhere. Thus we may only consider modes which are either symmetric or antisymmetric about the FCB to be valid, such that at BB2 they match onto the nonsingular series from the first big bang (BB1). These symmetry conditions can also be interpreted as a ``reflecting boundary condition'' at the FCB, since conformal time forms a double cover of cosmic time.

From these symmetry conditions, we arrive at having only a discrete set of comoving wave numbers, $k$, such that the allowed modes undergo the correct number of cycles between BB1 and BB2. This is analogous to an infinite potential well in which boundary conditions lead to quantized solutions with a particular set of wave numbers.  This allowed set of wave numbers has been analytically explored in \citet{Lasenby} for flat-$\Lambda$ radiation-dominated and matter-dominated universes and numerically found for a concordance $\Lambda$CDM universe in \citet{Deaglan}.

\begin{figure*}
    \centering
    \includegraphics{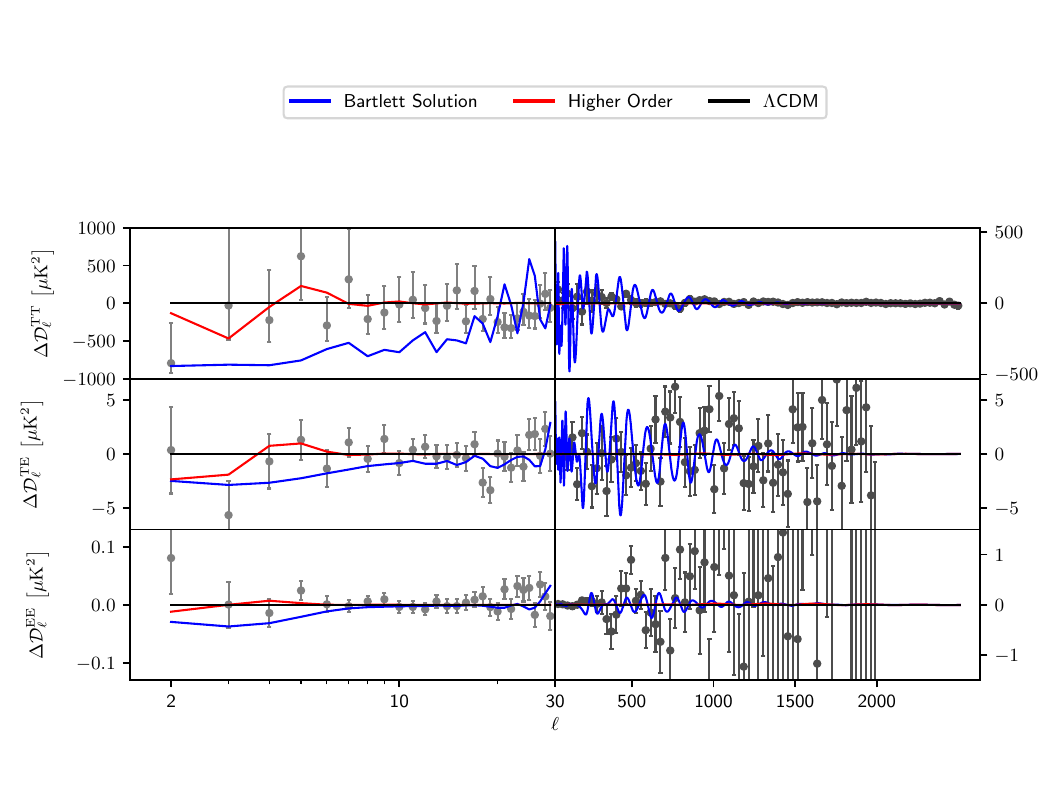}
    \caption{CMB power spectrum residuals between the quantized model calculated in this paper and the $\Lambda$CDM baseline. The corresponding curves for the Bartlett solution (\citet{Deaglan}) are also plotted for reference, as well as Planck residuals. The spectra produced by this new set of $k$ values appear to be more consistent with the data than the Bartlett solution $k$ values, but we seem to lose the interesting drop in power at ${\ell \approx 20}$ produced in the Bartlett solution. This figure was produced using an adapted version of CLASS~\cite{CLASS} used in~\cite{Deaglan}.}
    \label{fig:spectra}
\end{figure*}

Working with discrete comoving wave numbers gives rise to a different cosmic microwave background (CMB) power spectrum in comparison to the canonical calculation which uses a continuous set of $k$. We can, therefore, compute the predicted power spectra from the allowed wave numbers and compare these to observational data~\cite{PlanckCMB}, as shown in \cref{fig:spectra}. The specific sets of $k$ derived in previous work have been shown not to produce quantitatively good fits to current cosmological data, due to an unphysically large lowest allowed wave number, $k_0$; however, it has been demonstrated that a linearly spaced set of $k$ has the potential to provide significantly improved fits compared with the baseline concordance model of cosmology (\citet{Deaglan}). Moreover, they are capable of qualitatively reproducing some of the interesting low-multipole features of the CMB power spectrum, which have generated much discussion and various potential explanations over the years~\cite{Copi_2006, Copi_2010, Osti_2014, Gan_2018}.  Consequently, there are compelling reasons to investigate alternative quantized models which might predict these superior fits \emph{a priori}. 

\begin{figure*}
    \centering
    \includegraphics{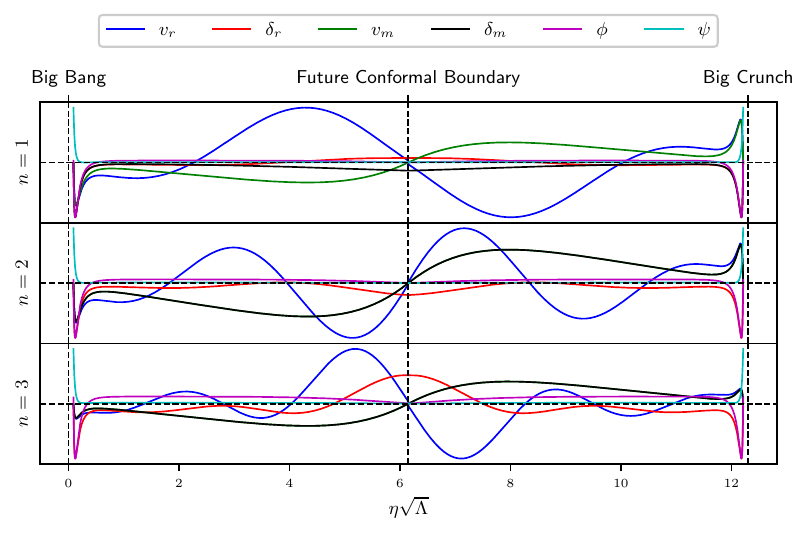}
    \caption{Base variable solutions are plotted for the first few allowed modes. The solutions become more oscillatory as $n$ increases.}
\label{fig:basevars}
\end{figure*}

\begin{figure*}
    \centering
    \includegraphics{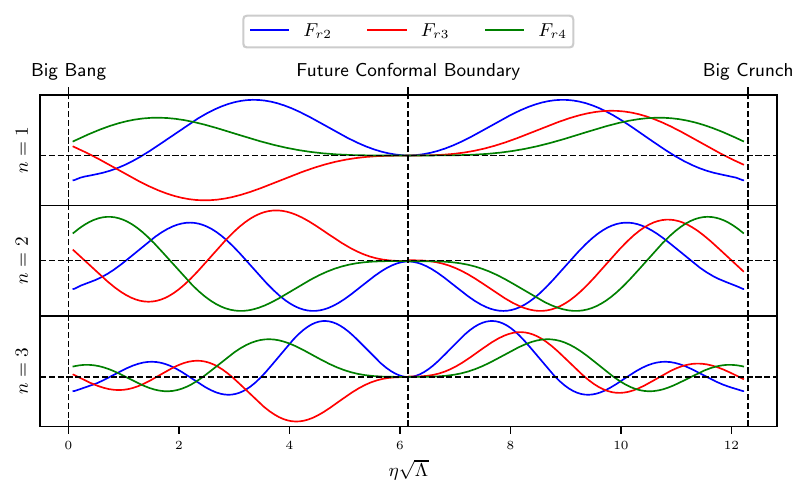}
    \caption{The first three anisotropic variables are plotted from the end of recombination for the first few allowed modes. We can see that these allowed modes have nonzero recombination values which are $n$-dependent.}
    \label{fig:highervars}
\end{figure*}

In \citet{Lasenby} and \citet{Deaglan} postrecombination photons are treated as a fluid with anisotropic stress (termed ``imperfect fluids''), which can be interpreted as including the first three terms in a photonic Boltzmann hierarchy.  
In this paper we build on this work by including the full photonic Boltzmann hierarchy. 
Making this extension requires a more sophisticated treatment of recombination modeling, whereby further free parameters are introduced via the exact values of higher order terms at the end of recombination; this enables us to have the correct number of degrees of freedom to satisfy the new quantization conditions required within this model and obtain a more accurate set of $k$ values. The first few allowed modes under this improved modeling of recombination are plotted for the base and some higher order variables in \cref{fig:basevars,fig:highervars}.

In Sec.~\ref{sec:level2}, we describe the background and perturbation equations we will be solving and present an overview of the work begun in \citet{Lasenby} and \citet{Deaglan}. We explain how to extend previous work to derive the new set of quantization conditions necessary for these higher order terms and describe the modeling of recombination that has been used here in Sec.~\ref{sec:level3}. In Sec.~\ref{sec:level4} the methods used to calculate the new set of allowed comoving wave numbers within this model are described, with the results and its implications discussed in Sec.~\ref{sec:level42}. Finally, conclusions are presented in Sec.~\ref{sec:level5}. 

Throughout this paper we work in units of ${8\pi G = c = \hbar = 1}$ and all overdots denote differentiation with respect to conformal time, unless otherwise explicitly stated. There are many symbols and subscripts used in the following sections, most of which follow the notation of \cite{Bert&Ma} and \cite{Lasenby, Deaglan}, so as a guide to the reader, we have summarized the key ones in Table \ref{tab:symbols}.

\begin{table}[]
    \centering
    \begin{tabular}{| c | c |}
    \hline
     Symbol &  Meaning \\ \hline\hline
      $k$ &  wave number of Fourier mode \\ \hline
     $a$   & background scale factor \\ \hline
    $s$  &  reciprocal scale factor\\ \hline
    $H_0$  &  present day Hubble parameter value\\ \hline
    $H_\infty$ & \multicolumn{1}{p{6cm}|}{\centering Hubble constant at FCB ($H_0\sqrt{\Omega_\Lambda}$)}\\ \hline
    $\Omega_i$  &  $i$th fluid's dimensionless density parameter\\ \hline
    $w_i$ & equation of state parameter for fluid $i$\\ \hline
     $\eta$   &  conformal time \\ \hline
     subscript  $r$  & radiation \\ \hline
    subscript $m$  & matter \\ \hline
    subscript $\Lambda$  & cosmological constant\\ \hline
     $\phi$ & Newtonian gauge potential\\ \hline
     $\psi$  & Newtonian gauge potential\\ \hline
     $\delta$  & perturbation to background density\\ \hline
    $v$   & peculiar velocity of density perturbation\\ \hline
    $\mathcal{H}$  &  conformal Hubble rate ($\frac{\dot{a}}{a})$ \\\hline
    $F_\ell$  & \multicolumn{1}{p{6cm}|}{\centering momentum-averaged Legendre components of perturbation to photon momentum distribution function}\\ \hline
    $G_\ell$  & photon polarization component \\ \hline
    $n_e$ & proper mean density of electrons \\ \hline
    $\sigma_T$ & Thomson scattering cross section\\ \hline
    superscript $\infty$ & quantity evaluated at FCB \\ \hline
    superscript $\ast$ & quantity evaluated at recombination \\ \hline
    \textbf{$x$}  & vector containing base variables\\ \hline
    $\textbf{y}_{2:4}$ & vector [$F_{r2}$, $F_{r3}$]\\ \hline
    $\textbf{y}_{4:}$ & \multicolumn{1}{p{6cm}|}{\centering vector containing all higher order variables apart from $F_{r2}$ and $F_{r3}$}\\ \hline
    $o_m$ & \multicolumn{1}{p{6cm}|}{\centering 'reduced matter parameter', defined as $H_\infty^3\frac{\Omega_m}{\Omega_\Lambda}$}\\ \hline
    $o_r$ & \multicolumn{1}{p{6cm}|}{\centering 'reduced radiation parameter', defined as $H_\infty^4\frac{\Omega_r}{\Omega_\Lambda}$}\\ \hline
    $K$ & dimensionless wave number, $\frac{k}{\sqrt{\Lambda}}$ \\ \hline
    \end{tabular}
    \caption{summarizing the symbols, subscripts and superscripts used throughout this paper.}
    \label{tab:symbols}
\end{table}

\section{\label{sec:level2} Theoretical Background}


\subsection{\label{sec:level21} Cosmological background equation}

For a homogeneous, isotropic and expanding universe, the most general metric is the Friedmann-Robertson-Walker (FRW) metric. Substituting this into the Einstein field equations and working within a perfect fluid approximation, we arrive at the background equation describing the evolution of such a universe:

\begin{equation}
    \Dot{s} ^2 = H_{0}^2 \sum_i \Omega_{i,0} |s|^{3(1 + w_i)},
\end{equation}
where ${s = \frac{1}{a}}$ is the reciprocal scale factor, $H_0$ is the present-day Hubble parameter value, $\Omega_i$ is the $i$th fluid's dimensionless density parameter and $w_i$ is the equation of state parameter for a given fluid $i$~\cite{Hobson, Deaglan}. Here, the subscripts $r$, $m$ and $\Lambda$ are used to denote radiation, matter and the cosmological constant respectively. We use modulus signs in this case as $s<0$ when analytically continued beyond the FCB.


\subsection{\label{sec:level22} Cosmological perturbation equations}

\subsubsection{\label{sec:level221} Perfect fluid approximation}

We must now include perturbations to this background in our equations. The notation used in this section and in the rest of the paper largely follows~\cite{Bert&Ma, Deaglan}, and we work in the conformal Newtonian gauge. 

If we assume that all components behave as perfect fluids, then it is possible to write the perturbation equations to linear order in matrix form, in a manner analogous to~\cite{Bamber}:

\begin{gather}
    \dot{\textbf{x}} = \label{eq:perf2} M(\eta)\textbf{x};\\
    \textbf{x} = 
    \begin{pmatrix}
    \phi, & \delta_r, & \delta_m, & v_r, & v_m
    \end{pmatrix}; \\
    \label{eq:perf3} M = \small{\begin{pmatrix}
        -\mathcal{H} & 0 & 0 & -2H_0^2 \Omega_{r} s^2 & -\frac{3}{2}H_0^2 \Omega_m |s| \\
        -4\mathcal{H} & 0 & 0 & \frac{4}{3} k^2 - 8H_0^2 \Omega_{r} s^2 & -6H_0^2 \Omega_m |s|\\
        -3\mathcal{H} & 0 & 0 & - 6H_0^2 \Omega_{r} s^2 & k^2 - \frac{9}{2}H_0^2 \Omega_m |s|\\
        -1 & -\frac{1}{4} & 0 & 0 & 0\\
        -1 & 0 & 0 & 0 & -\mathcal{H} \\
    \end{pmatrix}},
\end{gather}
where ${\mathcal{H} = \frac{\dot{a}}{a}}$ is the conformal Hubble rate, perturbations to the background densities are denoted by $\delta_i$, the peculiar velocities of these perturbations are denoted by $v_i$, and $\phi$ represents the Newtonian gauge potential~\cite{Dodelson}. 

\subsubsection{\label{sec:level222} Higher orders}

After recombination, given the lack of free electrons available to scatter and isotropize the radiation, the evolution of the photon distribution can be expressed using $F_\gamma(\vec{k}, \hat{n}, \tau)$, and $G_\gamma(\vec{k}, \hat{n}, \tau)$, the sum and difference, respectively,  of the phase space density of the two polarization components of linearly polarized photons~\cite{Bert&Ma}. Following~\citet{Lasenby} and \citet{Deaglan} we do not consider the $G_{r\ell}$ terms further due to these decoupling from the rest of the equations once our approximation of free-streaming postrecombination is made, but unlike in~\citet{Lasenby} and~\citet{Deaglan} we do not assume $F_{r\ell}$ to be zero for $\ell>2$.

The matrix representation of the perturbation equations within this Boltzmann hierarchy is given below, where we now redefine \textbf{x}  and $M(\eta)$ to include $\psi$, another conformal Newtonian gauge potential:

\begin{gather}
    \label{linear}\begin{pmatrix}
    \dot{\textbf{x}} \\ \dot{\textbf{y}}
    \end{pmatrix} = 
    \begin{pmatrix}
    M & N \\
    O & P \\
    \end{pmatrix}
    \begin{pmatrix}
    \textbf{x} \\ \textbf{y}
    \end{pmatrix};\\
    \label{xdef}\textbf{x} = 
    \begin{pmatrix}
    \phi, & \psi, & \delta_r, & \delta_m, & v_r, & v_m
    \end{pmatrix}; \\
     \label{ydef}\textbf{y} = 
    \begin{pmatrix}
    F_{r2}, & F_{r3},& \cdots
    \end{pmatrix};
    \end{gather}
    \begin{gather}
    M = \begin{psmallmatrix}
     0 & -\mathcal{H} & 0 & 0 & -2 H_0^2 \Omega_r s^2 & -\frac{3}{2} H_0^2 \Omega_m |s|\\
     0 & -\mathcal{H} & 0 & 0 & -\frac{2}{5} H_0^2 \Omega_r s^2 &  -\frac{3}{2} H_0^2 \Omega_m |s| \\
     0 & -4\mathcal{H} & 0 & 0 & -8 H_0^2 \Omega_r s^2 + \frac{4}{3}k^2 &  -6 H_0^2 \Omega_m |s| \\
     0 & -3\mathcal{H} & 0 & 0 & -6 H_0^2 \Omega_r s^2 & -\frac{9}{2} H_0^2 \Omega_m |s| + k^2\\
     0 & -1 &  -\frac{1}{4} & 0 & 0 & 0 \\
      0 & -1 & 0 & 0 & 0 & -\mathcal{H}\\
    \end{psmallmatrix}; \\
    N = 
    \begin{pmatrix}
    0 & 0 & \cdots \\
    \frac{6H_0^2 \Omega_r \mathcal{H} s^2}{k^2} & \frac{9}{5}\frac{H_0^2 \Omega_r s^2}{k} & \cdots \\
    0 & 0 & \cdots \\
    0 & 0 & \cdots \\
    \frac{1}{2} & 0 & \cdots \\
    0 & 0 & \cdots \\
    \end{pmatrix};
    \end{gather}
    \begin{gather}
    O = 
    \begin{pmatrix}
    0 & 0 & 0 & 0 & -\frac{8}{15} k^2 & 0 \\
    0 & 0 & 0 & 0 & 0 & 0 \\
    \vdots & \vdots & \vdots & \vdots & \vdots & \vdots \\
    \end{pmatrix}; \\
    P = 
    \begin{pmatrix}
    -\frac{9}{10} \frac{n_e \sigma_T}{s} & -\frac{3}{5}k & \cdots\\
    \frac{3k}{7} & - \frac{n_e \sigma_T}{s} & \cdots \\
    \vdots & \vdots & \ddots \\
    \end{pmatrix},
\end{gather}where $n_e$ is the proper mean density of electrons and $\sigma_T$ is the Thomson scattering cross section. 

In general, the higher order derivative terms may be written as

\begin{equation}\label{eq:thomson}
    \dot{F}_{r\ell} = \frac{k}{2\ell + 1} \big[\ell F_{r(\ell-1)} - (\ell +1)F_{r(\ell+1)}\big] - \frac{n_e \sigma_T}{|s|} F_{r\ell},
\end{equation}
where, following~\cite{Bert&Ma}, when we truncate the equations at some $\ell_\textrm{max}$ we set
\begin{equation}\label{truncation}
    \dot{F}_{r\ell_\textrm{max}} = kF_{r(\ell_\textrm{max}-1)} - \bigg(\frac{\ell_\textrm{max} + 1}{\eta} + \frac{n_e \sigma_T}{|s|} \bigg) F_{r\ell_\textrm{max}}.
\end{equation}

The Newtonian gauge potentials can also be written explicitly in terms of the other variables, which will be of use when calculating power series expansions about the future conformal boundary later:

\begin{equation}\label{phieqn}
    \phi = - \frac{3H_0^2}{2k^2} \bigg( \Omega_m |s| (\delta_m + 3\frac{\dot{s}}{s} v_m) + \Omega_r s^2 (\delta_r + 4\frac{\dot{s}}{s} v_r) \bigg); \\
\end{equation}

\begin{equation}\label{psieqn}
    \psi = \phi - \frac{3H_0^2 \Omega_r}{k^2} s^2 F_{r2}.    
\end{equation}

In this paper we do not solve for $\phi$ and $\psi$ explicitly, but this will be discussed further in Sec.~\ref{sec:level411}. 


\subsection{\label{sec:level23} Imposing symmetry for Bartlett case}

Below we summarize the argument first discussed in~\citet{Lasenby} and then reformulated in~\citet{Deaglan} for imposing the quantization condition ${v_r^\infty = 0}$, where the superscript denotes that \textbf{$v_r$} is evaluated at the FCB.

In order to prevent divergence of our solutions at BB2, the first genuine singularity after ${\eta=0}$, we must impose either symmetry or antisymmetry about the FCB on all of the perturbations. We can see the consequences of this more clearly if we write out the power series expansion of $\phi$ about the FCB, to linear order:

\begin{eqnarray}
    \phi = -\frac{3}{2k^2} \Big[\textrm{sign}(\Delta\eta) \times H_\infty^3 \frac{\Omega_m}{\Omega_\Lambda} (\delta_m^\infty + 3\dot{v}_m^\infty) 
    \nonumber \\
    + 4H_\infty^4 \frac{\Omega_r}{\Omega_\Lambda} v_r^\infty \Big] (\Delta\eta) + O(\Delta\eta^3),
\end{eqnarray}
where ${H_\infty = H_0 \sqrt{\Omega_\Lambda}}$ is the Hubble constant evaluated at the FCB and $\Delta\eta$ is the conformal time difference to the actual FCB, defined to be positive before the FCB.

If $v_r^\infty$ is nonzero, then in order to impose either symmetry or antisymmetry on the above expression we must require that the term depending on the sign of $\Delta\eta$ is the same either side of the FCB. However, it can be shown that this means we are unable to apply the symmetry condition to $v_m$ and a contradiction is reached. As such, all valid solutions within this scheme must obey ${v_r^\infty = 0}$ (such that $\phi$ is always symmetric about the FCB). This is only satisfied for a discrete set of $k$ values.


\section{\label{sec:level3} Theory}

In the following section we calculate the new quantization conditions at the future conformal boundary which need to be satisfied by the perturbation variables. The key result is given in Eq. ~\ref{conditions}. 

\subsection{\label{sec:level31} Deriving FCB quantization conditions}

We may extend the above argument by considering the power series expansion of $\phi$ up to third order about the FCB when we include higher order terms of the Boltzmann hierarchy. This may be written as

\begin{widetext}

\begin{eqnarray}\label{phiseries}
     \phi = -\frac{3}{2k^2}\Bigg[\bigg(\textrm{sign}(\Delta\eta) \times H_\infty^3 \frac{\Omega_m}{\Omega_\Lambda}(\delta_m^\infty + 3\dot{v}_m^\infty) + 4H_\infty^4 \frac{\Omega_r}{\Omega_\Lambda} v_r^\infty \bigg)\Delta\eta  
     + \bigg (2H_\infty^4 \frac{\Omega_r}{\Omega_\Lambda} F_{r2}^\infty \bigg) \Delta\eta^2 \nonumber \\
   + \bigg(\textrm{sign}(\Delta\eta) \times \frac{1}{2} H_\infty^3 \frac{\Omega_m}{\Omega_\Lambda} k^2 \dot{v}_m^\infty  + H_\infty^4 \frac{\Omega_r}{\Omega_\Lambda} (\frac{2}{15}k^2 v_r^\infty - \frac{3}{5} k F_{r3}^\infty \bigg) \Delta\eta^3 \Bigg] + O(\Delta\eta^4).
\end{eqnarray}

\end{widetext}

The key thing here is that if a variable is to be symmetric about the FCB, then the coefficients of odd powers of $\Delta\eta$ in its power series must swap sign either side of the FCB, such that the product of the coefficient and $\Delta\eta^n$ (where $n$ is odd) keeps the same sign on both sides of the boundary. Similarly, its coefficients of even powers of $\Delta\eta$ must be sign-independent either side of the FCB. The converse must be true for antisymmetric variables. 

Let us consider the third term in the $\phi$ series; in general we will not have $\dot{v}_m^\infty$ equal to zero. But it is shown in~\citet{Deaglan} that enforcing $v_r^\infty = 0$ results in $\dot{v}_m^\infty$ being continuous across the FCB and thus keeping the same sign. Hence, this first term of the third-order coefficient keeps $\phi$ symmetric about the FCB, as required. In order to continue this symmetry we must therefore enforce that the sign-independent term in this expression be zero at the FCB. Since we already have the condition that ${v_r^\infty = 0}$ from above, this is equivalent to requiring that ${F_{r3}^\infty = 0}$. We note here that considering the $\phi$ expansion alone leads to no constraints being placed on $F_{r2}^\infty$ since the sign-dependent term cancels in the coefficient of $\Delta\eta^2$.

After we have enforced the conditions ${v_r^\infty = F_{r3}^\infty = 0}$, we may write the power series expansion of $F_{r2}$ as

\begin{equation}
\begin{aligned}
     F_{r2} = F_{r2}^\infty + k^2\Big[\frac{1}{15}\delta_r^\infty - \frac{11}{42}F_{r2}^\infty + \frac{6}{35}F_{r4}^\infty\Big]\Delta\eta^2 \\ + O(\Delta\eta^3).
\end{aligned}
\end{equation}

Remembering that we require all the variables to be continuous across the FCB, we can see that the effect of $\delta_r^\infty$ being in general nonzero is that $F_{r2}$ is forced to be symmetric about the FCB. In fact, it can be shown (see Appendix \ref{AppA}) that for every $F_{r\ell}$ the value of $\delta_r^\infty$ appears in the coefficient of $\Delta\eta^\ell$ in the power series. This forces all variables with odd $\ell$ to be antisymmetric and those with even $\ell$ to be symmetric about the FCB. 

This has profound consequences for the quantization conditions. If all even modes are required to be symmetric, their first-order derivatives at the FCB must vanish so that the coefficient of $\Delta\eta$ in their power series will be zero. This is because the first order coefficients can be written entirely in terms of FCB values of higher order variables, which, if not equal to zero at the FCB, are required by continuity to keep the same sign either side of it. But if they keep the same sign, then the overall first-order term will swap sign either side of the FCB, breaking the symmetry requirement. Because of the way the odd and even modes are coupled in the equations, the condition of the first order derivative of even modes disappearing is equivalent to requiring that ${F_{r\ell}^{\infty} = 0}$ for all odd $\ell$ (see Appendix \ref{AppB} for further detail). 

Finally, let us consider the coefficient of $\Delta\eta^3$ in $F_{r2}$'s power series. The third order term is where we start to include the Thomson scattering term in the power series. Since $n_e$ refers to the proper electron density, it is proportional to $|s|^3$. Thus the whole Thomson scattering term from \cref{eq:thomson} is proportional to $s^2$ and for clarity we shall write it as $Bs^2F_{r\ell}$, where the constant $B$ encompasses both the electron density and the Thomson scattering cross section. In this case the third-order coefficient of $F_{r2}$'s series can now be expressed in the form

\begin{equation}
\begin{aligned}
    F_{r2}^{(3)} & = \frac{8}{45}k^2 \phi^{(1)} - \frac{3}{15}kF_{r3}^{(2)} - BH_\infty^2F_{r2}^\infty,
\end{aligned}
\end{equation}
where $\phi^{(1)}$ refers to the coefficient of $\Delta\eta$ in $\phi$'s power series and $F_{r3}^{(2)}$ refers to the coefficient of $\Delta\eta^2$ in $F_{r3}$'s power series. 

Since $F_{r3}$ is antisymmetric about the FCB, it follows that $F_{r3}^{(2)}$ at the FCB must either be zero or depend on the sign of $\Delta\eta$ (in fact, it is zero since all odd $F_{r\ell}^\infty$ are zero). We recall from \cref{phiseries} that $\phi^{(1)}$ also depends on the sign of $\Delta\eta$. These both therefore keep $F_{r2}$ symmetric about the FCB, as required by the arguments above. However, it can be seen that $F_{r2}^\infty$ must be set to zero if we are to keep this symmetry, since otherwise this term is multiplied by $\Delta\eta^3$, which changes sign either side of the FCB. 

The third order terms for any even mode, $n$, may be written similarly to above, purely in terms of second order odd $F_{rl}$ terms and a term proportional to $F_{rn}^\infty$, which can be seen from Eq. (\ref{eq:thomson}). Following the same argument we are forced to set all ${F_{r\ell}^\infty = 0}$ for even $\ell$.

We thus arrive at our final set of quantization conditions which are required to enforce the correct symmetry on our equations:

\begin{eqnarray}\label{conditions}
\begin{aligned}
     v_r^\infty & = 0; \\ 
     F_{rl}^\infty & = 0 \hspace{3mm} \textrm{for} \hspace{3mm} \ell \geq 2.
\end{aligned}
\end{eqnarray}
Conversely~\citet{Lasenby} and~\citet{Deaglan} assume that $v_r^\infty =0$,  $F_{rl}^\ast = 0$, where $\ast$ refers to the time of recombination --- a quite different set of boundary conditions!

\subsection{\label{sec:level32} Modeling of recombination}

Until recombination we work within the perfect fluid approximation. This means that all higher order terms are set to zero and thus are equal to zero at the start of recombination. In \citet{Lasenby} and \citet{Deaglan}, recombination was assumed to be instantaneous: i.e. since we require the variables to be continuous at all conformal times, the higher order terms must be zero at the end of recombination too, at which point we begin solving our Boltzmann hierarchy. If all the higher order terms are initialized to zero at the end of recombination, there will only be one free parameter, the wave number, $k$, in our set of equations and there is no guarantee of a solution to Eq.~\ref{conditions}. 

If, instead of assuming that recombination occurs instantaneously, we consider that it happens over a small finite time so that the higher order modes have time to grow from zero at the start of recombination to some finite value by the end of recombination, then this introduces additional free parameters into our equations in the form of the exact recombination values of the higher order terms. We will further assume that the effect of this growth on the base variables is negligible, such that the base variables values do not change significantly during recombination.

With this more sophisticated modeling of recombination, we have the same number of free parameters as quantization conditions, where the final free parameter is $k$. Thus we will arrive at a discrete set of $k$ values for which the quantization conditions are satisfied, as in~\citet{Lasenby} and~\citet{Deaglan}. 

\section{\label{sec:level4} Methods}

Throughout this paper we use the perfect fluid equations from Sec.~\ref{sec:level221} before recombination and then assume free-streaming (${n_e = 0}$) afterwards; in reality, the Thomson scattering term will be small, hence why we neglect it in this section and assume free-streaming, but it is not actually zero and so must be included when considering boundary conditions. We may calculate the recombination values of base variables by integrating the perfect fluid perturbation equations from ${\eta=0}$ to recombination, using adiabatic initial conditions~\citep{Lasenby,Deaglan}. 

\subsection{\label{sec:level411} General approach}

Since the perturbations can be described by a system of linear differential equations, as in \cref{linear}, their solutions form a vector space~\cite{lang}. This means we are able to encode a linear mapping between solutions at times $\eta_0$ and $\eta_1$ using a transfer matrix:

\begin{eqnarray}
x(\eta_1) = U(\eta_1, \eta_0)x(\eta_0).
\end{eqnarray}

For the case of cosmological perturbations, we may only solve for this transfer matrix numerically, by integrating the perturbation equations between $\eta_0$ and $\eta_1$ with initial conditions [1,0,0,...], [0,1,0,...] etc. to find each column. 

In this case, we are in the somewhat unusual situation of solving a differential equation where some of the variables are specified at one boundary, the time of recombination, and some at the FCB. Since the system is linear, we can solve this without ``shooting'' methods but, in theory, we still have to integrate either one way or the other in all of the variables in order to obtain the transfer matrix. It makes sense to start the integration at the FCB since we know much more of the variables at this boundary.

So if we now take ${\eta_1=\eta_\ast}$ to be the conformal time at recombination and ${\eta_0=\eta_\infty}$ to be at the FCB, we can relate the perturbations at these times similarly using a matrix. Using the same notation for base and anisotropic variables as in \cref{xdef,ydef}), we may write this explicitly as

\begin{equation} \label{simplebackint}
\begin{pmatrix}
\textbf{x}^\ast \\ \textbf{y}^\ast
\end{pmatrix} = 
\begin{pmatrix}
A & B \\
C & D \\
\end{pmatrix}
\begin{pmatrix}
\textbf{x}^\infty \\ \textbf{y}^\infty
\end{pmatrix},
\end{equation}
where the superscript, $\ast$, refers to values at recombination and $A,B,C,D$ are submatrices within the transformation matrix. The values of $\textbf{x}^\ast$ are known from integrating the perfect fluid equations from ${\eta=0}$ to recombination, in the same manner as in~\citet{Lasenby} and~\citet{Deaglan}. Our modeling of recombination means that the values $\textbf{y}^\ast$ are treated as free parameters and are unknown. On the right-hand side we know what values $\textbf{y}^\infty$ should take for the allowed modes, from Eq.~\ref{conditions}, but, apart from $v_r^\infty, \phi^\infty$ and $\psi^\infty$, the values of $\textbf{x}^\infty$ are unknowns. 

This last point is worth examining more closely: when considering the simultaneous equations produced by the base variables, we see that there will be six equations (from six base variables). But it has also been shown, for example in \cref{phiseries}, that $\phi^\infty$ and $\psi^\infty$ are always equal to zero regardless of $k$. This means that there are really only four unknown variables within these six equations. The key thing to note here is \cref{phieqn,psieqn}, which show that both $\phi$ and $\psi$ can be determined entirely by the higher order and other base variables and are hence not truly independent variables themselves. So upon closer inspection we see that there is no issue with having six equations describing four unknowns; we may simply discard the top two superfluous equations when solving for the FCB values of the other variables, and we can use \cref{phieqn,psieqn} to check for consistency.  

Going back to \cref{simplebackint}, let us explicitly write out the top row:

\begin{equation}
    \textbf{x}^\ast = A\textbf{x}^\infty + B\textbf{y}^\infty.
\end{equation}

Given that we know we want ${\textbf{y}^\infty = \textbf{0}}$ in the allowed modes and we know $\textbf{x}^\ast$, for each $k$ we can solve for $\textbf{x}^\infty$ and then choose the $k$ values for which ${v_r^\infty = 0}$. This is the general approach we will use to find the allowed wave numbers. At this stage, we are not interested in the bottom row of \cref{simplebackint} as this does not help to solve for $v_r^\infty$ but can be later used to find the recombination values of higher order terms. 

\subsection{\label{sec:level412} Integrating backwards from the FCB}

In order to obtain the transfer matrix between FCB and recombination values, we must integrate backwards from the FCB using initial conditions of $[1,0,0,...], [0,1,0,...], [0,0,1,...]$ etc. to find each column of the matrix. This is possible in theory but in practice starting our integration at the exact FCB is highly numerically unstable. Fortunately, it becomes markedly more stable when we start at small deviations from the FCB. Here we have chosen to begin the integration at ${\Delta\eta = 10^{-3}}$ before the FCB as this keeps the error in $v_r^\infty$ relatively small while still being close enough to the FCB that we can make efficient use of power series expansions to initialize the variables.

In order to maintain reasonable errors in $v_r^\infty$, we must specify the initial conditions up to at least third order in their power series. We must calculate the expansions explicitly for all variables in the hierarchy up to $\ell = 3$: for higher order variables we may set the initial condition to zero, where we have used the fact that enforcing the FCB quantization condition leads to the first nonzero term in the power series of $F_{r\ell}$ being proportional to $\Delta\eta^\ell$. 

In order to distinguish those higher order variables which need to be specified at the FCB using power series expansions and those which can simply be set to zero, we now split $\textbf{y}$ so that $\textbf{y}_{2:4}$ denotes the vector $[F_{r2}, F_{r3}]$, and the $\ell \geq 4$ terms are contained within $\textbf{y}_{4:}$. Using the superscript~$'$ to denote the values of variables at our chosen start point, ${\eta^{'} = \eta_\textrm{FCB} - \Delta\eta}$, we may now rewrite \cref{simplebackint} as
\begin{equation}\label{fullbackint}
    \begin{pmatrix}
    \textbf{x}^\ast \\ \textbf{y}_{2:4}^\ast \\ \textbf{y}_{4:}^\ast
    \end{pmatrix} = 
    \begin{pmatrix}
    A & B_{2:4,2:4} & B_{4:,4:} \\
    C & D_{2:4,2:4} & D_{4:,4:} \\
    \end{pmatrix}
    \begin{pmatrix}
    \textbf{x}^{'} \\ \textbf{y}_{2:4}^{'} \\ \textbf{y}_{4:}^{'}
    \end{pmatrix},
\end{equation}
where $\textbf{y}_{4:}^{'}$ is $\textbf{0}$ and 
\begin{equation}\label{appendixeqn}
    \begin{pmatrix}
    \textbf{x}^{'} \\ \textbf{y}_{2:4}^{'}
    \end{pmatrix} = 
    \begin{pmatrix}
    X_1 \\
    X_2  \\
    \end{pmatrix}
    \textbf{X}^\infty.
\end{equation}

In the above, $\textbf{X}^\infty$ represents the vector $[\delta_r^\infty, \delta_m^\infty, v_r^\infty, \dot{v}_m^\infty]$. The submatrices $X_1$ and $X_2$ are given by
\begin{widetext}

\newcommand\scalemath[2]{\scalebox{#1}{\mbox{\ensuremath{\displaystyle #2}}}}

\begin{gather} 
X_1 = 
    \scalemath{0.8}{
    \begin{pmatrix}
    0 & -\frac{3}{2k^2}o_m\Delta\eta & o_r(\frac{6}{k^2}\Delta\eta + \frac{1}{5}\Delta\eta^3) & o_m(-\frac{9}{2k^2}\Delta\eta - \frac{3}{4}\Delta\eta^3) \\
    0 & -\frac{3}{2k^2}o_m\Delta\eta & o_r(\frac{6}{k^2}\Delta\eta + \frac{1}{5}\Delta\eta^3) & o_m(-\frac{9}{2k^2}\Delta\eta - \frac{3}{4}\Delta\eta^3) \\
    1 - \frac{1}{6}k^2 \Delta\eta^2 & o_m(-\frac{6}{k^2}\Delta\eta + \frac{2}{3}\Delta\eta^3) & -\frac{4}{3k^2}(k^4 - 18o_r)\Delta\eta - (\frac{28}{15}o_r - \frac{2}{15}k^4)\Delta\eta^3 & o_m(-\frac{18}{k^2}\Delta\eta - \Delta\eta^3) \\
    0 & 1 + o_m(-\frac{9}{2k^2}\Delta\eta + \frac{1}{2}\Delta\eta^3) & o_r(\frac{18}{k^2}\Delta\eta - \frac{7}{5}\Delta\eta^3) & \frac{1}{2}k^2\Delta\eta^2 - o_m(\frac{27}{2k^2}\Delta\eta + \frac{3}{4}\Delta\eta^3) \\
    \frac{1}{4}\Delta\eta - \frac{1}{40}k^2\Delta\eta^3 & -\frac{3}{2k^2}o_m\Delta\eta^2 & 1 + \frac{1}{k^2}(6o_r - \frac{3}{10}k^4\Delta\eta^2 & -\frac{9}{2k^2}o_m\Delta\eta^2 \\
    0 & -\frac{3}{2k^2}o_m\Delta\eta^2 & \frac{6}{k^2}o_r\Delta\eta^2 & -\Delta\eta - \frac{9}{2k^2}o_m\Delta\eta^2
    \end{pmatrix}
    }; \\
    X_2 = 
    \scalemath{0.9}{
    \begin{pmatrix}
    \frac{1}{15}k^2\Delta\eta^2 & -\frac{4}{15}o_m\Delta\eta^3 & \frac{8}{15}k^2\Delta\eta - \frac{16}{15}(\frac{1}{14}k^4 - o_r)\Delta\eta^3 & -\frac{4}{5}o_m\Delta\eta^3 \\
    -\frac{1}{105}k^3\Delta\eta^3 & 0 & -\frac{4}{35}k^3\Delta\eta^2 & 0 \\
    \end{pmatrix}
    },
\end{gather}
\end{widetext}where $o_m = H_\infty^3 \frac{\Omega_m}{\Omega_\Lambda}$ and $o_r = H_\infty^4 \frac{\Omega_r}{\Omega_\Lambda}$ are reduced matter and radiation parameters,
and simply contain the coefficients of the power series expansions for the base, $F_{r2}$ and $F_{r3}$ variables.

Writing out the top row of \cref{fullbackint} as

\begin{equation}\label{solveforvrinf}
    \textbf{x}^\ast = (AX_1 + B_{2:4,2:4}X_2)\textbf{X}^\infty + B_{4:,4:}\textbf{0},
\end{equation}
we see a major simplification to the problem: given that the matrix $B_{4:,4:}$ is always going to be multiplied by the zero vector, there is no need to find its components explicitly. We are thus able to greatly reduce our computation time by only having to perform eight integrations to determine the first eight columns of each transfer matrix (six for the base variables and two more for $F_{r2}$ and $F_{r3}$), and we can set the remaining columns to zero. This is especially useful given that we are unable to use traditional time-saving approximations used in Boltzmann codes, such as estimating the higher order terms using spherical Bessel functions~\cite{CLASS}, when we get close to the FCB.   

\subsection{\label{sec:level413} Implementation of code}

Throughout the code we work with dimensionless units for the comoving wave number by writing ${k = K\sqrt{\Lambda}}$. We also choose units such that $\sqrt{\Lambda}$ is unity, so we may write the present-day Hubble value as ${H_0 = \frac{1}{3\Omega_\Lambda}}$. All relevant $\Lambda$CDM parameters are taken from the Planck best-fit values, given in the posterior samples TT+lowE~\cite{Planck2018}. The rescaled values are given in \cref{tab:planckvals}.

\begin{table}[]
    \centering
    \begin{tabular}{c c}
    \hline\hline
    Parameter & \multicolumn{1}{p{3cm}}{\centering TT+lowE \\ 68\% limits} \\ \hline
     $\Omega_{\Lambda}$   &  0.679 \\
     $\Omega_{m}$   & 0.321 \\
     $\Omega_{r}$ & $9.24 \times 10^{-5}$ \\
     $\textit{H}_0$ & 0.701 \\
      $z^{\ast}$ & 1090.30 \\
      \hline
    \end{tabular}
    \caption{Planck best-fit values from~\cite{Planck2018}, after being rescaled to new units of ${8\pi G = c = \hbar = \Lambda = 1}$.}
    \label{tab:planckvals}
\end{table}

For each value of $K$ between 0 and 20, the background and perturbation equations were integrated, using the necessary initial conditions, from the FCB to recombination using LSODA from \texttt{scipy.integrate.solve\_ivp}. Once the transfer matrix was obtained in this way, \cref{solveforvrinf} was set up and solved using \texttt{numpy.linalg.solve} to find the value of $v_r^\infty$. This was then plotted against $K$ and the allowed modes were determined by finding where this curve intersected the $x$-axis. 

\section{\label{sec:level42} Results and Discussion}

\subsection{\label{sec:level421} Final allowed modes}

\begin{figure}
    \centering
    \includegraphics{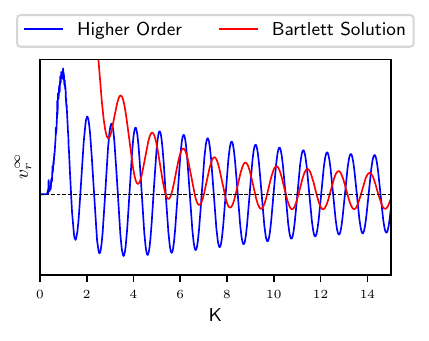}
    \caption{Converged values of $v_r^\infty$, solved using the method described in Sec.~\ref{sec:level4}, are plotted as a function of the dimensionless comoving wave number $K$. The equivalent curve for the Bartlett solution~\cite{Deaglan} is also plotted for reference.}
    \label{fig:finalmodes}
\end{figure}

\begin{figure}
    \centering
    \includegraphics{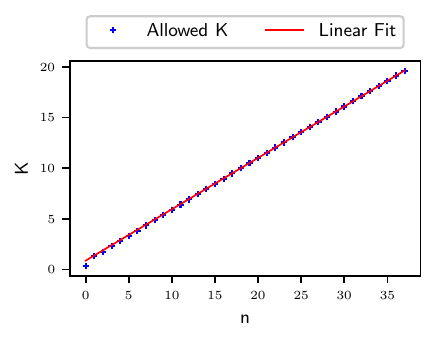}
    \caption{The first 38 allowed wave numbers have been explicitly calculated and plotted. It is clear from the excellent agreement of the linear fit that the spacing between the allowed $K$ values quickly settles down to a constant. }
    \label{fig:linear}
\end{figure}

The final, converged, graph of $v_r^\infty$ against $K$ is presented in \cref{fig:finalmodes}, with $\ell_{max}=70$, a choice which will be justified in the following section; the corresponding curve for the Bartlett solution is included for a direct comparison. Qualitatively, we can see that the allowed modes, when higher order terms of the Boltzmann hierarchy are included, are more closely spaced than in the Bartlett solution. As in~\citet{Lasenby} and ~\citet{Deaglan}, although the spacing of the modes is initially non-linear, it settles down to linear fairly quickly, as shown in \cref{fig:linear}. 

This linear spacing, as well as the lowest allowed wave number, is

\begin{eqnarray}
    \label{k0} k_0 &= 0.309 \sqrt{\Lambda} = 9.93 \times 10^{-5} \textrm{Mpc}^{-1}\\
    \label{k1} k_1 &=1.328 \sqrt{\Lambda} = 4.27 \times 10^{-3} \textrm{Mpc}^{-1}\\
   \label{delk} \Delta k &= 0.507\sqrt{\Lambda} = 1.63 \times 10^{-4} \textrm{Mpc}^{-1}.
\end{eqnarray}
For reference, the equivalent values for the Bartlett imperfect fluids solution are~\cite{Deaglan}:
\begin{eqnarray}
    k_0^{\textrm{Bartlett}} = 0.042\sqrt{\Lambda} = 1.34 \times 10^{-5} \textrm{Mpc}^{-1}\\
    k_1^{\textrm{Bartlett}} = 5.39\sqrt{\Lambda} = 1.73 \times 10^{-3} \textrm{Mpc}^{-1}\\
    \Delta k^{\textrm{Bartlett}} = 0.657\sqrt{\Lambda} = 2.11 \times 10^{-4} \textrm{Mpc}^{-1}.
\end{eqnarray}

A key difference between the two curves in \cref{fig:finalmodes} is that the second allowed mode occurs at a much smaller $K$ value (${K = 1.33}$, as opposed to $5.39$) and the ``missing modes'' from the Bartlett solution are reintroduced. In fact, these modes are reintroduced even for the case where we consider the $\ell=2$ terms, if we take $F_{r2}$ as a free parameter. The other difference is that the new spacing between the allowed modes is smaller. The first few allowed modes are plotted in \cref{fig:basevars,fig:highervars}.

\subsection{\label{sec:level422} Convergence of allowed modes}

We can examine the degree of convergence of solutions by taking several different values of $\ell_\textrm{max}$, the order at which the Boltzmann hierarchy is truncated, and plot the curves of $v_r^\infty$ against $K$ for each of these to see how changing $\ell_\textrm{max}$ shifts the allowed modes. This is demonstrated in \cref{fig:reflection}, which will be discussed further in the following section. 

In fact, the results show a remarkable convergence of the curves, even at relatively low values of $\ell_\textrm{max}$, indicating that the transition from the Bartlett case is due to just a few critical low-$\ell$ modes.

\subsection{\label{sec:level423} Truncation and artificial reflection}

We must truncate our Boltzmann hierarchy at some $\ell_\textrm{max}$, and a smaller $\ell_\textrm{max}$ leads to a less computationally-expensive code. However, truncating too early leads to some unexpected behavior of our graphs, and the regular oscillating behavior of $v_r^\infty$ in $K$-space is lost. This is likely due to artificial reflection of power from $\ell_\textrm{max}$ back to lower multipoles, as suggested in~\cite{CLASS}. Although the alternative truncation scheme quoted in \cref{truncation}, proposed in~\cite{Bert&Ma}, is designed to minimize this, significant levels of unphysical reflection cannot be avoided for late times. The value of $\ell_\textrm{max}$ should thus scale with $k\eta$. 

\begin{figure}
    \centering
    \includegraphics{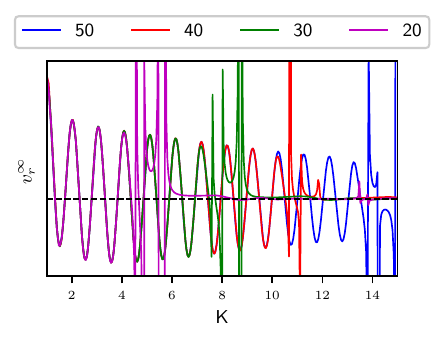}
    \caption{$v_r^\infty$ is plotted as a function of $K$ for various values of $\ell_\textrm{max}$, the order at which our Boltzmann hierarchy is truncated. For values of ${\ell_\textrm{max}/K \lesssim 4}$, the curves begin to misbehave, potentially due to unphysical reflection as warned in~\cite{CLASS}.}
    \label{fig:reflection}
\end{figure}

In order to demonstrate the effect of truncation, graphs of $v_r^\infty$ against $K$ have been plotted for ${\ell_\textrm{max} = 20, 30, 40}$ and $50$. As can be seen in \cref{fig:reflection}, the graphs begin to deviate from their regular structure at ${\ell_\textrm{max}/K \approx 4}$. It is interesting to note, however, that before the graphs begin to diverge, they still give the same allowed modes; this again reinforces that the transition to these new allowed modes is likely due to just a few critical low-$\ell$ modes.

\subsection{\label{sec:level424} The choice of root finding variable}

It should be noted that, although in this paper we have chosen to search for zeros in $v_r^\infty$ to determine the allowed modes, this decision is largely arbitrary. Since it does not matter in which order we solve the quantization conditions, we could in theory set $v_r^\infty$ to zero initially and then search for zeros in any of the other higher order multipoles. These will all lead to the same allowed modes, and all have the same issue of artificial reflection of power for too small $\ell_\textrm{max}$ so, numerically speaking, there is no clear advantage to any. One advantage of looking for zeros in $v_r^\infty$ is that it is more directly comparable to the work in~\citet{Lasenby} and~\citet{Deaglan}. 

We could also start by setting $v_r^\infty$ to zero as well, use three of the four base variable equations to solve for the remaining three FCB values and then use the fourth equation to match onto the recombination value of the fourth variable. Again, this is of a similar stability to the other methods provided we do not truncate the equations too early. 

\subsection{\label{sec:level425} Discussion of $k_0$ and $\Delta k$}

We may use the calculated values of $k_0$ and $\Delta k$ from ~\cref{k0,delk} to compute a $C_\ell$ power spectrum and compare this with observational data. However, in~\citet{Deaglan} the general class of linearly quantized models was assessed so we may use these for an initial comparison of our values. 

In Figs. 5 and 8 of~\citet{Deaglan}, likelihood plots showing the difference in quality of fit between the linearly quantized models and $\Lambda$CDM models are given, as functions of $k_0$ and $\Delta k$. Qualitatively, it appears that the linearly quantized model calculated in this project gives roughly the same quality of fit as the $\Lambda$CDM model. However, it is important to note that the models in~\citet{Deaglan} were examined using a profile likelihood analysis. \citet{Thomas} have shown that the conclusions may change if investigated using a Bayesian approach, although more work is required to test the full suite of predictions against matter power spectrum constraints.

\Cref{fig:spectra} shows the CMB power spectrum residuals between the quantized model found in this project and the $\Lambda$CDM baseline, produced using the code from~\citet{Deaglan} which is an adapted version of CLASS~\cite{CLASS}. It can be seen that the resulting spectra in this case appear to be more consistent with cosmological data than the Bartlett solution, but it is interesting to see that we lose some of the noteworthy low-$\ell$ features of the latter, such as the dip in power at ${20 \lesssim \ell \lesssim 30}$. The spectra shown in \cref{fig:spectra} should be interpreted with some caveats as for a fair comparison we ought to explicitly calculate within a discrete model the $C_\ell$ power spectrum arising from this quantization; indeed, this leaves the possibility that the quantized model may even provide a better fit to the data than the $\Lambda$CDM baseline. 

\section{\label{sec:level5} Conclusion}

We have extended the results of~\citet{Lasenby} and~\citet{Deaglan} to include higher order terms of the Boltzmann hierarchy in calculating the quantized spectrum of comoving wave numbers for a palindromic universe containing radiation, dark matter and a cosmological constant. We derived a new set of quantization conditions needed to impose the correct symmetry on the solutions to the cosmological perturbation equations, such that they do not diverge at the singularity after the future conformal boundary. A more sophisticated modeling of the evolution of perturbations through recombination has been employed to enable a consistent set of solutions to these quantization conditions. Using this, the discrete set of comoving wave numbers satisfying these conditions were calculated by exploiting the linear nature of the equations. 

The lowest permissible wave number within this model was calculated to be ${k_0 = 9.93 \times 10^{-5} \textrm{Mpc}^{-1}}$ and allowed modes are separated by a linear spacing of ${\Delta k = 1.63 \times 10^{-4} \textrm{Mpc}^{-1}}$. An initial comparison of these values to the general class of linearly quantized models considered in~\citet{Deaglan} indicate that the spectra produced from this quantization are fairly consistent with observational data. However, the $C_\ell$ power spectrum must be explicitly calculated within a discrete model in order to fairly assess the quality of agreement of this model with observed data.

The spacing of these allowed modes ceases to change upon addition of further higher order terms at a relatively low value of $\ell_\textrm{max}$, indicating that the transition from the Bartlett solution occurs due to just a few critical low-$\ell$ modes. 

In addition to computing the $C_\ell$ spectrum, further work could explore a more detailed modeling of recombination, as well as including the Thomson scattering term in the integration of variables. A Bayesian analysis could also be performed on the general class of linearly quantized models in order to obtain updated fits for ${(k_0, \Delta k)}$ values. One of the main limitations of the methods described in this paper is the long run-time of the code to solve the Boltzmann hierarchy at large $k$ values, due to the highly oscillatory nature of the solutions. More efficient methods have recently been developed to solve differential equations with rapidly oscillating solutions, such as the RKWKB method for one-dimensional systems~\cite{Agocs} and Magnus expansion based methods for higher-dimensional ones~\cite{Bamber}, but more work needs to be done before the use of these can be extended up to high $\ell$ values. A recent paper~\cite{altformalism} has explored an alternative formulation of the perturbation equations, whereby the Boltzmann hierarchy is replaced by just two integral equations describing the photon intensity quadrupole and the linear-polarization quadrupole, and such techniques could prove more numerically suitable when executing a full pipeline confronting these models of the Universe against the latest cosmological data.

\acknowledgements

We thank Deaglan Bartlett for helpful conversations about his work on this subject, and Carola Zanoletti for her input. M.P. thanks the Cavendish Laboratory for the Part III Project opportunity. W.H. is supported by a Royal Society University Research Fellowship.

\appendix

\section{\label{AppA}}

Below we argue that $\delta_r^\infty$ always appears in the coefficient of $\Delta\eta^\ell$ in the power series expansions of higher order variables about the FCB. This is used in Sec.~\ref{sec:level31} to argue that all variables with odd $\ell$ are forced to be antisymmetric about the FCB, and all variables with even $\ell$ are forced to be symmetric.

The power series expansion about the FCB for a higher order variable may be written in general as:

\begin{eqnarray}\label{eq:genpowerseries}
    F_{r\ell} = F_{r\ell}^\infty + \dot{F}_{r\ell}^\infty\Delta\eta + \frac{1}{2}\ddot{F}_{r\ell}^\infty\Delta\eta^2 + \ldots\nonumber \\+ \frac{1}{n!}F_{r\ell}^{(n)\infty}\Delta\eta^n + \ldots
\end{eqnarray}

Now, $\dot{F}_{r\ell}$ will depend on $F_{r(\ell-1)}$, $F_{r\ell}$ and $F_{r(\ell+1)}$ only [from Eq. (\ref{eq:thomson})]. Let us consider the coefficient of $\Delta\eta^\ell$ in the above power series:

$F_{r\ell}^{(\ell)}$ will depend on $F_{r(\ell-1)}^{(\ell-1)}$, $F_{r(\ell)}^{(\ell-1)}$ and $F_{r(\ell+1)}^{(\ell-1)}$ only. But, for example, $F_{r(\ell-1)}^{(\ell-1)}$ will depend on $F_{r(\ell-2)}^{(\ell-2)}$, $F_{r(\ell-1)}^{(\ell-2)}$ and $F_{r(\ell)}^{(\ell-2)}$. If we keep following this through until we express the coefficient in terms of just the higher order variables themselves (as opposed to derivatives of them), then eventually $\dot{v}_r^\infty$ will be used, which depends on $\delta_r^\infty$. This is the smallest coefficient for which $\delta_r^\infty$ appears as for the coefficient of $\Delta\eta^{(\ell-1)}$, if we express this just in terms of the higher order variables, the first term will be proportional to $v_r^\infty$, and for the coefficient of $\Delta\eta^{(\ell-2)}$ the first term will be proportional to $F_{r2}^\infty$ and so on. 

\section{\label{AppB}}

Here we show that requiring the first order derivatives of even $\ell$ terms to disappear is equivalent to requiring that $F_{r\ell} = 0$ for all odd $\ell$. The first order derivative of an even $\ell$ variable may be written as:

\begin{equation}
    \dot{F}_{r\ell} = \frac{k}{2\ell + 1} \big[\ell F_{r(\ell-1)} - (\ell +1)F_{r(\ell+1)}\big] - Bs^2 F_{r\ell},
\end{equation}
from Eq. (\ref{eq:thomson}) and from Sec.~\ref{sec:level31} where we write the Thomson term as being proportional to $s^2$. However, in the power series solution, $s$ must be written as a power series expansion about the FCB too; $s$ can be written as $H_\infty\Delta\eta + ...$ so in the first order power series term, $\dot{F}_{r\ell}\Delta\eta$, the Thomson term is in fact proportional to $\Delta\eta^3$ and is thus not first order. So to first order in $\Delta\eta$, we only have $\frac{k}{2\ell + 1} \big[\ell F_{r(\ell-1)} - (\ell +1)F_{r(\ell+1)}\big]$ as the coefficient to $\Delta\eta$. So requiring that this coefficient disappears is equivalent to requiring that $\ell F_{r(\ell-1)} = (\ell+1) F_{r(\ell+1)}$ for all even $\ell$.  

Let us start from $\ell=4$: since we have already enforced $F_{r3}^\infty = 0$ earlier on in the argument in Sec.~\ref{sec:level31}, this means that $F_{r5}^\infty$ is now also forced to be zero. Similarly, for $\ell=6$, since $F_{r5}^\infty$ has been forced to be zero, $F_{r7}^\infty$ is now also forced to be zero, and so on. Thus, for all odd $\ell$ we require that $F_{r\ell}^\infty$ be zero. 

\bibliography{references}

\end{document}